\documentclass{aa}

\usepackage{graphics} 

\usepackage[T1]{fontenc}
\usepackage{times}

\begin{document}

\thesaurus{11.06.2; 11.09.1 Phoenix; 11.12.1; 11.19.5; 11.19.6}

\title{Stellar populations in the Phoenix dwarf galaxy.
\thanks{Based on data collected at the European Southern Observatory,
         La Silla, Chile, Proposal N.57.A-0788}
}
\subtitle{}

\author {Enrico V. Held  \inst{1} \and Ivo Saviane \inst{2} 
\and Yazan Momany \inst{2} }
\offprints {E. V. Held}

\institute{
Osservatorio Astronomico di Padova, 
vicolo dell'Osservatorio 5, I--35122 Padova, Italy
\and
Universit\`a di Padova, Dipartimento di Astronomia, 
vicolo dell'Osservatorio 5, I--35122 Padova, Italy
}

\date {}

\maketitle

\begin{abstract}

We have obtained deep CCD photometry in the $B$, $V$, and $I$\ bands of  
Phoenix, a galaxy considered a transition case between dwarf spheroidal 
(dSph) and dwarf irregular (dI) galaxies. 
A comparison  of our data with the giant branches of Galactic globular 
clusters gives a mean metal abundance  [Fe/H]$=-1.81\pm0.10$\ dex. 
The presence of an intrinsic color dispersion in the upper red giant branch 
(RGB) suggests an abundance range of about 0.5 dex, although a range in age 
may also affect the RGB width. 
The color-magnitude diagram ({\sc cmd}) of Phoenix reveals for the first time a 
horizontal branch (HB) predominantly red yet moderately extended to the blue, 
similar to those of Leo~II  or And~I,  at $V \approx 23.8$.  
The detection of a relatively blue HB  indicates the presence of 
a significant population with age comparable to that of old halo Galactic 
globular clusters.  As in other dwarf spheroidals, this HB morphology 
in a metal-poor system indicates a mild ``second parameter''  effect. 
The mean level of the HB has been used to derive a true distance modulus 
$23.21\pm0.08$, in good agreement with the distance modulus 
$23.04\pm0.07$\  estimated from the well defined cutoff of the red giant 
branch at $I\approx23.1$. 
This confirms the correct identification of the RGB tip. 
We also find a radial gradient in the Phoenix HB morphology, as measured 
by an increasing ratio of  blue HB stars to red giant stars 
in the outskirts of the galaxy. 
The color-magnitude diagrams show a small number of stars above the tip of 
the RGB,  well in excess over field contamination, that most likely are 
asymptotic giant branch (AGB)  stars belonging to an intermediate age 
population. Their number indicates that the fraction of intermediate age (3 
to 10 Gyr) population in Phoenix is approximately 30--40\%. 
A young stellar population is definitely present in Phoenix, 
consistent with a  star formation episode started at least 
0.6 Gyr ago, up to $1\times10^{8}$\  yr ago. 
Both young stars and AGB stars are centrally concentrated, which  indicates 
that recent star formation preferentially occurred in the inner galaxy 
regions.  
In many respects, including an extended star formation history and even the 
presence of a modest amount of neutral hydrogen, Phoenix appears  not 
dissimilar from dwarf spheroidal galaxies in the Local Group. 

\end{abstract}

\keywords{Galaxies: fundamental parameters -- 
{\bf Galaxies: individual: Phoenix} 
-- Local Group -- Galaxies: stellar content }

\section{Introduction}
\label{s_intro}

Photometric studies of dwarf spheroidals, and in particular of their
color-magnitude diagrams, have shown a variety of star 
formation histories, ranging from the globular cluster-like diagrams of Draco or 
Tucana   (Grillmair et al. \cite{gril+97}; Saviane {{et al.}\ } \cite{savi+96}, hereafter 
Paper~I) to the considerable intermediate age
populations  of Carina (Smecker-Hane {{et al.}\ } \cite{smec+94}) and
Fornax (Beauchamp {{et al.}\ } \cite{beau+95}; Stetson {{et al.}\ } \cite{stet+98}; 
Saviane \& Held \cite{savi+held98}), 
with a range of intermediate cases for which we refer the reader to 
the recent reviews of  Da Costa (\cite{daco98}) and Mateo (\cite{mate98}).

The dwarf galaxy Phoenix represented  a particularly interesting case in this respect 
since it is currently forming stars on top of a seemingly old, metal-poor population.
Ortolani \& Gratton (\cite{orto+grat88}, hereafter OG88) obtained $B, V$\ 
CCD photometry in a $2^\prime\times3^\prime$\  field centered on 
the blue star association first described by Canterna \& Flower 
(\cite{cant+flow77}).     Their color-magnitude diagrams show  bright blue stars likely 
belonging to a 
$10^8$\ yr-old recent burst of star 
formation, involving $\sim10^4$\ {$M_{\odot}$}, superposed onto a 
predominantly old ($>10^{10}$\ yr) metal-poor stellar population.  
A comparison with metal-poor Galactic globular 
clusters led OG88 to estimate a distance modulus $(m-M)\approx23-23.5$, 
thus locating Phoenix well within the Local Group. 
OG88 also found a $(B-V)$ color distribution on the RGB somewhat 
broader than accounted for by instrumental errors, the intrinsic color 
dispersion corresponding to an abundance range of about 0.6 dex. 
A further study of the resolved stellar populations of Phoenix by van de Rydt 
et al. (\cite{vdry+91}, hereafter VDK91), yielded a slightly shorter distance 
modulus of $(m-M)_0=23.1$\  and a  metal abundance  [Fe/H]$=-2.0$. 
The  integrated luminosity, $M_V\approx-9.9$, is similar to that of Carina.  
These photometric studies together with limited spectroscopic data of 
Da Costa (\cite{daco94}), also indicated the presence of low-luminosity carbon stars  
consistent with an intermediate age population older than 8--10 Gyr.  

Both studies concluded that Phoenix may belong to  an intermediate class between dI's 
and dSph's, a view supported by the detection of neutral  hydrogen towards 
the optical galaxy
at a heliocentric radial  velocity of 56 {km s$^{\rm -1}$}\ 
(Carignan et al. \cite{cari+91}). 
More recently,  Young \& Lo (\cite{youn+lo97})  obtained 
VLA maps of  {\hbox{H{\sc i}}}\ in a region around Phoenix at a resolution 
$\sim2^\prime$. Their maps show  an {\hbox{H{\sc i}}}\ cloud  (``cloud B'') at about 
$8^\prime-10^\prime$\ from the optical center, at a heliocentric 
velocity of $50-60$\  {km s$^{\rm -1}$}, which is identified with the emission 
detected by Carignan et al. (\cite{cari+91}).  Its velocity dispersion is 
low, only $\sim 3$ {km s$^{\rm -1}$}. In addition, that study revealed a new {\hbox{H{\sc i}}}\ feature 
peaked at $\sim -23$\ {km s$^{\rm -1}$}, about $5^\prime$\ southwest of the 
optical galaxy (``cloud A''), whose curved shape suggests a physical link 
with the galaxy (see the {\hbox{H{\sc i}}}\ contour maps in their Fig.~17).   
However, it is conceivable that  both clouds are associated with the 
complex {\hbox{H{\sc i}}}\  velocity structure of the Magellanic Stream.

The presence of young stars along with neutral gas in an essentially old 
galaxy may provide the clue to understanding the role of multiple bursts of 
star formation in the evolution of dwarf spheroidals (and possibly of their 
more massive counterpart, the dwarf ellipticals). 
For this reason, we have undertaken a study of the old, young, and 
intermediate age populations of Phoenix.
A first task was establishing  whether Phoenix 
harbors a significant populations of metal-poor stars  of age comparable to 
Galactic globular clusters, in which case a  blue HB is expected. 
Indeed, we have been able to detect for the first time a moderately blue HB 
at $V \approx 23.7$. A further aim was a quantitative estimate of the 
contribution of young and  intermediate age stars to the stellar populations of Phoenix, 
in particular to understand if the recent burst of star formation 
represents an isolated episode after a long quiescent period. 

The plan of the paper is as follows. Section~\ref{s_obsred} presents our 
new observations in $B$, $V$, and $I$\  of the Phoenix dwarf over a
 $6\farcm9\times6\farcm9$\ field. Careful analysis of the instrumental errors 
and foreground/background contamination, essential  to estimate 
the contribution of the young and intermediate age stellar 
populations, is also presented. 
In Sect.~\ref{s_cmd} we describe the main features of the 
color-magnitude diagrams, and outline the spatial distribution of stars of 
different ages.  In particular, we  introduce our first detection of the 
horizontal branch of  Phoenix. 
 The basic properties of the galaxy are derived in Sect.~\ref{s_analy}.
There we  provide a new distance estimate based both on the $I$\ 
magnitude of the luminosity of the RGB tip, and on the mean $V$\  
magnitude of  horizontal branch stars. 
The wide photometric baseline employed in this study allows us to 
re-address the problem of an intrinsic color range and abundance dispersion of 
the red giants. 
The stellar populations of the Phoenix dwarf  are discussed in 
Sect.~\ref{s_discu}, where in particular we estimate the contribution of young 
and intermediate age stars to the galaxy luminosity, and map the distribution 
of the star-forming regions across the galaxy. 
The main results of this paper are summarized in Sect.~\ref{s_summa} along 
with concluding remarks.  

\section{Observations and data reduction}
\label{s_obsred}

Observations of  a single $6\farcm 9 \times 6 \farcm 9$ field centered on 
Phoenix were obtained on September 7 and 8, 1996 using EFOSC2 at the 
ESO/MPI~2.2m telescope.  The detector was a thinned  Loral CCD with 
$2048\times2048$  pixels of 0\farcs26. The images were trimmed to 
$1600\times1600$ pixels during readout to avoid the vignetted edges of the 
image.   Fast readout, appropriate for broad band imaging,  yielded a 
read-out noise of 7.8~e$^{-}$/pixel  and a conversion factor 
of 1.34~e$^{-}$/ADU.  
A control field located at a few degrees from the galaxy was also observed 
to estimate the  foreground and galaxy background contamination. 
Sky conditions were photometric on the second night of the run.  The seeing 
was  quite variable, so that only the best pairs of images in each filter were 
selected for processing. 
Table~\ref{t_obsjou} lists the images used in this study, including  the 
comparison field exposures.  Images with poorer seeing  (not listed 
here)  were used for calibration purposes, namely to evaluate the zero 
point stability. 
 
 \begin{table}[h]
 \caption{The journal of observations \label{t_obsjou}}
\begin{flushleft}\begin{tabular}{lccccc}
 \noalign{\smallskip}\hline\noalign{\smallskip}
 \multicolumn{1}{c}{ID} &
 \multicolumn{1}{c}{Night} &
 \multicolumn{1}{c}{Filter} &
 \multicolumn{1}{c}{$t_{\rm exp}$[s]} &
 \multicolumn{1}{c}{airmass} &
 \multicolumn{1}{c}{FWHM$\arcsec$} \\
 \noalign{\smallskip}\hline\noalign{\smallskip}
 %
 
  Ph/b5 & 2 &  B & 2400 & 1.04 &  1.4 \\ 
  Ph/b6 & 2 &  B & 2400 & 1.04 &  1.3 \\ 
 
 
  Ph/i2 & 2 &  I & 2400 & 1.06 & 1.2  \\ 
  Ph/i3 & 2 &  I & 1800 & 1.10 &  1.3 \\ 
 
 
  Ph/v1 & 1 &  V & 1500 & 1.29 & 1.5  \\ 
  Ph/v2 & 1 &  V & 1500 & 1.21 &  1.4 \\ 
 
 
    &       &    &    &       &        \\
  bck/b1  & 1 & B  & 2400 & 1.04 & 1.3 \\
  bck/b2  & 1 & B  & 2400 & 1.06 & 1.3 \\
  bck/i1  & 1 & I  & 2400 & 1.07 & 1.3 \\
  bck/i2  & 1 & I  & 2400 & 1.04 & 1.3 \\
  bck/v1  & 1 & V  & 1500 & 1.15 & 1.7 \\
  bck/v2  & 1 & V  & 1500 & 1.11 & 1.7 \\
  \noalign{\smallskip}\hline\end{tabular}\end{flushleft}\end{table}

 \begin{figure}
 %
 \resizebox{\hsize}{!}{\includegraphics{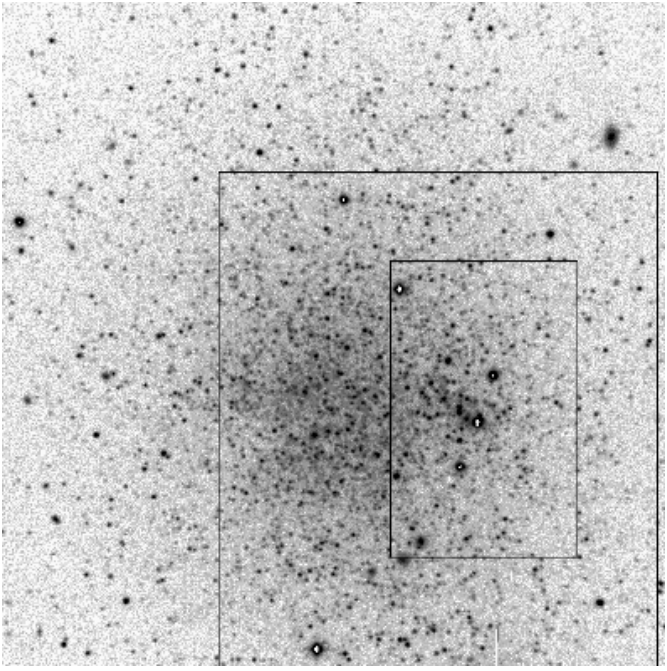}}
 \caption[]{A central field ($5\farcm3 \times 5\farcm3$) of the coadded $V$ 
 image of Phoenix.  North is to the top and East to the left.  
 The outlines indicate the regions studied by Ortolani \& Gratton 
 (\cite{orto+grat88}; inner box)  
 and van de Rydt et al. (\cite{vdry+91}; outer box)
 }
 \label{f_magmap}
 \end{figure}

Image pre-processing was carried out with the {\sc eso midas}
package  in the standard way. 
First, all images were  cleaned using a map of the bad features of the CCD. 
After checking  the bias stability, a master bias 
was subtracted from all images. 
Several daylight dome flats and twilight flat-fields were obtained in  each 
filter and used to construct dome and sky master flat-fields. The sky flats 
were preferred for correcting the scientific frames, because of the higher 
signal-to-noise  ratio and better color match, though the dome flats were 
almost 
as good. 
The images were then registered to a common 
reference frame and coadded.  The size of the Point Spread Function (PSF) 
of the averaged images shows no degradation with respect to the individual 
frames.  As an example,  our $V$\ sum image is shown in 
Fig.~\ref{f_magmap}.

Stellar photometry was performed using {\sc daophot ii} and {\sc allstar} 
(Stetson \cite{stet87}). 
The PSF was iteratively constructed  from a starting list of about 100 
(relatively bright) reference stars, uniformly distributed across the CCD. 
After fitting a preliminary PSF to the sum images, all reference stars with 
faint neighbors were culled out of the catalog by careful visual inspection 
of the star-subtracted images. A better point spread function was then 
obtained from the selected  list of PSF stars.  
Different PSF shapes were tested by checking the residuals after 
subtraction.  The best fitting  was obtained using a Moffat function with 
$\beta=1.5$\  and quadratic radial dependence. The subtraction residuals 
were  generally satisfactory due to the good sampling of stellar images. 
Photometry  was then obtained by running {\sc daophot} and {\sc allstar} 
twice on the sum images.

 %
 \begin{table}[h]
 \caption{ The photometric errors ($1\sigma$) determined from the
 artificial star experiments \label{t_taberr}}
\begin{flushleft}\begin{tabular}{r|rrr}
 \noalign{\smallskip}\hline\noalign{\smallskip}
 \multicolumn{1}{c}{bin} &
 \multicolumn{1}{c}{$\sigma_V$} &
 \multicolumn{1}{c}{$\sigma_B$} &
 \multicolumn{1}{c}{$\sigma_I$} \\
 
 \noalign{\smallskip}\hline\noalign{\smallskip}
 %
 %
 %
  16.25 & ~$\cdots$~ & ~$\cdots$~ &  0.004 \\
  16.75  & ~$\cdots$~ & ~$\cdots$~ &  0.014 \\
  17.25  & ~$\cdots$~ & ~$\cdots$~ &  0.012 \\
  17.75  & ~$\cdots$~ & ~$\cdots$~ &   0.013 \\
  18.25  & ~$\cdots$~ & ~$\cdots$~ &  0.015 \\
  18.75  & ~$\cdots$~ & ~$\cdots$~ &  0.015 \\
  19.25 & 0.007 & 0.025 & 0.021   \\ 
  19.75 & 0.009 & 0.006 &  0.023  \\ 
  20.25 & 0.014 & 0.010 &  0.031  \\ 
  20.75 & 0.019 & 0.013 &   0.043 \\ 
  21.25 & 0.024 & 0.021 &  0.061 \\ 
  21.75 & 0.032 & 0.028 &   0.087 \\ 
  22.25 & 0.041 & 0.031 &   0.116 \\ 
  22.75 & 0.059 & 0.041 &  0.124 \\
  23.25 & 0.079 & 0.056 &  0.145 \\ 
  23.75 & 0.163 & 0.078 &  0.101  \\ 
  24.25 & 0.105 & 0.087 &  ~$\cdots$~  \\ 
  24.75 & 0.111 & 0.148 &  ~$\cdots$~ \\
 \noalign{\smallskip}\hline\end{tabular}\end{flushleft}\end{table}

The photometric errors  and the degree of completeness of our data were 
estimated by extensive artificial star experiments.  We used a technique in 
which artificial stars added to each image  are randomly  scattered  about 
the vertices of  a  fixed grid or ``lattice''. This prevents overlapping of 
simulated stars (``self-crowding''). 
Lists of input stars 
(typically $\sim900$) were created for the $V$ image, with randomly
distributed coordinates and $V$ magnitudes uniformly distributed  within 
the limits of the Phoenix {\sc cmd}. 
Random ($B-V$) and ($V-I$) colors were employed, so that the artificial star 
colors spanned the entire useful intervals in the Phoenix color-magnitude 
diagrams. We made  30 experiments per filter, corresponding to $\simeq 27000$ stars 
in 
each bandpass. The frames containing the simulated stars were reduced 
exactly in the same way as the real images.  The master catalog  of star 
experiments contains the artificial stars retrieved  in at least one filter, ready 
for  calibration as the real photometry catalogs.

Table~\ref{t_taberr} lists the photometric errors obtained in each bandpass 
separately. The first column gives the center of the magnitude bin in $B$,
$V$, and $I$. The photometric errors were obtained by  grouping the stars 
in magnitude bins and calculating the differences  $\Delta m = m_{\rm out} - 
m_{\rm in}$\  between the retrieved and input magnitudes.  The standard 
deviations are those of a Gaussian fitted to the observed distribution of 
$\Delta m$. 
The $\Delta m$\  are typically 50\% 
larger than the {\sc daophot} errors.  
The completeness was calculated in a two-dimensional way by dividing each 
color-magnitude diagram in cells with size 0.2 mag in magnitude 
and 0.5 mag in color. Contours of equal completeness, smoothed over 
$3\times3$\ cells, are superimposed to the  color-magnitude diagrams.

Observations  of standard stars from Landolt (\cite{land92}) were 
used to calibrate the photometry. A value of $0.030$\ s was assumed  for
the shutter delay time, which implies  a correction  of 
0.003 mag  for the standard stars. 
 The raw magnitudes were normalized to 1 s exposure time and zero 
airmass: 
\begin{equation}
\label{e_eqnorm}
m' = m_{\rm ap} + 2.5\, \log (t_{\rm exp} + \Delta\,t) - k_{\lambda}\,X
\end{equation}
where $m_{\rm ap}$ are the instrumental magnitudes measured in circular
apertures of radius $R = 6\farcs6$\ (close to the photoelectric aperture 
employed by Landolt \cite{land92}), $\Delta\,t$ is the shutter delay,  and $X$
is the airmass. The adopted mean extinction coefficients for La Silla are 
 $k_B = 0.235$, $k_V = 0.135$, and $k_I = 0.048$.
A fit of the normalized instrumental magnitudes to the magnitudes of the 
standard stars in the Landolt's (\cite{land92}) fields gives the following 
relations:
\begin{eqnarray} \label{e_calibra}
B  &=& b' + 0.121 \, (B-V) + 24.777   \\
V  &=& v' + 0.0425 \, (B-V) + 25.145   \\
V  &=& v' + 0.0381 \, (V-I) + 25.145   \\
I   &=& i'  - 0.0253 \, (V-I) + 24.049, 
\end{eqnarray}

where the coefficients were derived from the standard stars observed on the 
second night. 
The r.m.s. scatter of the residuals of the fit (0.007, 0.008, and 0.008 mag in 
$B$, $V$, and $I$ respectively) was assumed to represent our calibration 
uncertainties.

Before applying these calibrating relations to our photometry, 
 the instrumental profile-fitting magnitudes (measured on the sum 
frames)  were converted to the scale of  the standard star 
measurements.  To this purpose,  a sample of bright isolated objects  were 
selected in all  individual images, all  neighboring faint stars subtracted, and 
the cleaned  frames were used to  measure magnitudes through circular 
digital apertures with radii $r_{\rm ap}=6\farcs6$ (the same as for standard 
stars).  Aperture corrections were then calculated for each image 
as the  median of  5 to 8   independent corrections in each filter.
The r.m.s. scatter about the median zero  point is 0.015 in $B$, 0.011 in $V$, 
and 0.012 in $I$. 
This uncertainty includes not only the errors on aperture measurements and 
corrections, but also the effects of extinction variations during the nights. 
Following aperture correction, 
calibrated magnitudes were obtained by applying the calibrating 
relations  in the form of a linear system solved  by the 
Kramer method (a procedure equivalent to the more usual iterative 
method).  The total zero-point uncertainties, obtained by quadratic sum of  
the aperture correction  and  calibration errors, 
are 0.017, 0.014, and 0.014 mag in $B$, $V$, and $I$\ respectively.

 \begin{figure}
 \resizebox{\hsize}{!}{\includegraphics{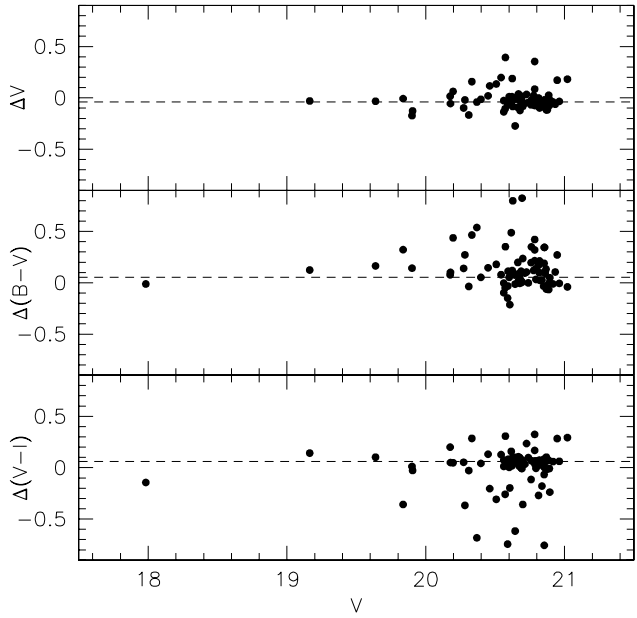}}
 \caption[]{Comparison of the present data with the photometry of
 van de Rydt et al. (\cite{vdry+91}). 
 The differences in $V$ (top panel), $B-V$
 (mid panel) and $V-I$ (bottom panel) are plotted for stars
 brighter than $V=21$.  The dashed lines represent the median 
 zero-point differences (this paper $-$\ VDK91) }
 \label{f_comp_vdr}
 \end{figure}

Figure~\ref{f_comp_vdr} shows a comparison of our photometry with the 
results of  van de Rydt {{et al.}\ } (\cite{vdry+91}) for stars brighter than $V=21$.  
We obtain the following median magnitude and color differences 
(this study $-$\ VDK91):  $\Delta V = -0.036 \pm 0.083$, 
$\Delta (B-V) = 0.091 \pm 0.116$, $\Delta (B-I) = 0.108 \pm 0.151$,  and
$\Delta (V-I) = 0.060\pm 0.101$.   The errors are standard deviations of the 
differences, following a $3\sigma$\ rejection of the outliers. 
Van de Rydt {{et al.}\ } (\cite{vdry+91}) report 
the result of their  comparison with the data of Ortolani \& Gratton 
(\cite{orto+grat88}). 
They found mean differences, in the sense VDK91 $-$\ OG88, 
of $\Delta V = -0.014 \pm 0.080$\  and $\Delta (B-V) =  -0.082 \pm 0.100$.
 Thus our ($B-V$) colors are somewhat redder than those of VDK91 and 
consistent with the results of OG88.  Our slightly brighter $V$\ magnitude 
scale is  explained  by the larger  reference aperture  (indeed, using the OG88 
aperture we would have obtained  a zero point dimmer by  $0.03\pm0.01$\ mag). 

\section{The color-magnitude diagrams }
\label{s_cmd}

 \begin{figure}
 \hbox{
 \resizebox{\hsize}{!}{\includegraphics{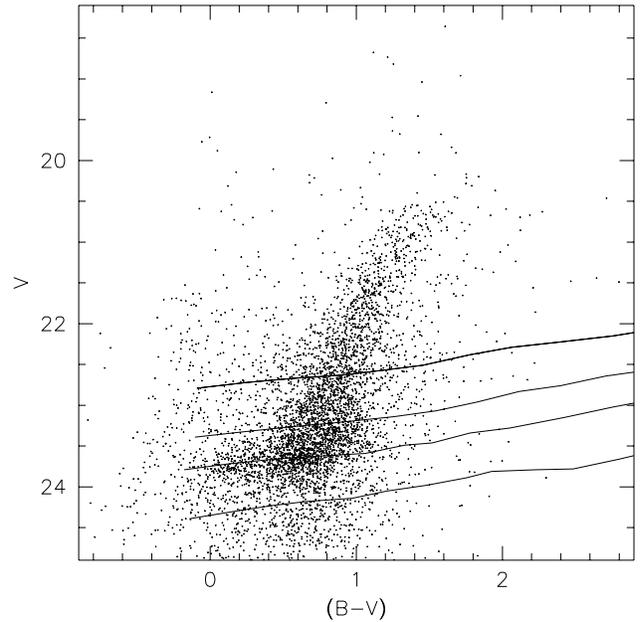}}
 }
 \caption[]{
 The $V$, $(B-V)$\ color-magnitude diagram of the Phoenix  dwarf galaxy, 
 from our master photometric catalog. No selection has been applied to the 
 data. 
 Also shown are the $10$\%,  $30\%$,  $50\%$, and $70\%$ isocompleteness 
 contours
 }
 \label{f_cmdbvv}
 \end{figure}

 \begin{figure*}
 \hbox{
 \resizebox{\hsize}{!}{\includegraphics{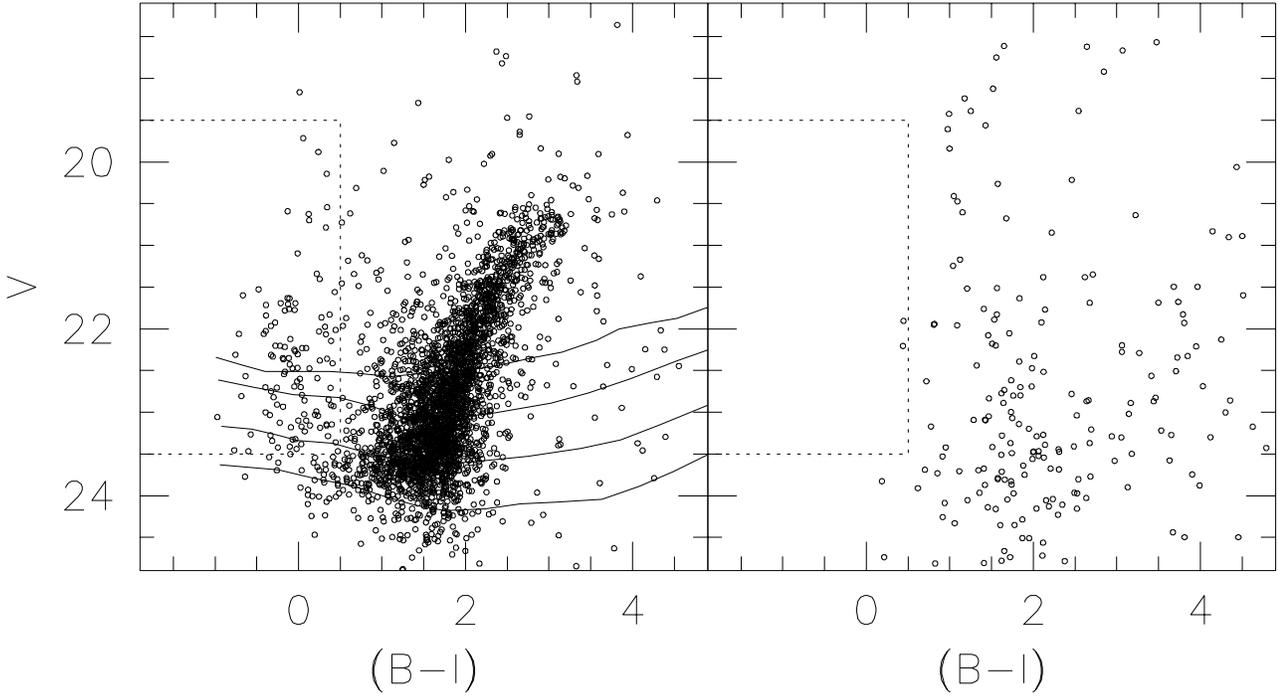}}
 }
 \caption[]{The $V$,  $(B-I)$  color-magnitude diagram including all stars 
 in the field of Phoenix ({\em left panel}). Superimposed are the 10\%, 30\%, 50\%,  
 and 70\% isocompleteness levels.  A dotted box marks the region chosen to represent 
 the blue sequence of  young stars.  The 
 {\em right panel} shows a similar diagram for stars in the comparison field
  } 
 \label{f_manycmd}
 \end{figure*}

Figure~\ref{f_cmdbvv} shows the $V$, $(B-V)$ color-magnitude diagram of 
Phoenix.  The most notable feature  is our detection of the horizontal branch 
(HB) of Phoenix at  $V\sim23.7$, discernible  up  to  
$(B-V)=0.0$.  An  extended blue tail, similar to that of metal-poor Galactic 
globular clusters, is not evident.  The HB is found at a magnitude level where 
our {\sc cmd}\  is significantly incomplete and photometric errors are quite 
severe, so deeper data are certainly important.  The morphology of the HB  
will be analyzed in more detail in Sect.~\ref{s_discu}.  
Figure~\ref{f_cmdbvv} also displays a red giant branch (RGB) sharply cut at
$V \sim 20.5$, and a blue sequence extending to luminosities 
brighter than the tip of the RGB, representing the young star population 
detected by OG88 and VDK91. 
This blue star sequence, partially overlapping  the HB region, is also 
noticeable in other c-m diagrams. Figure~\ref{f_manycmd} 
shows the color-magnitude diagram of Phoenix using the wider $(B-I)$\ 
color baseline.  The trend of the isocompleteness contours  
in Fig.~\ref{f_manycmd} illustrates  the relatively high incompleteness 
of  $I$-band  photometry in the case of blue stars. 
The {\sc cmd}\  of Phoenix is  compared with the similar {\sc cmd}\  of the control 
field in the right panel of Fig.~\ref{f_manycmd}. A  comparison of the two 
diagrams shows that the foreground and background  counts do not 
represent a major source of contamination for  the most interesting  regions 
of the {\sc cmd}.  
There are no foreground or background objects bluer than 
$(B-I) = 0.5$, so that  we are confident that all stars bluer than this limit 
belong to a young stellar population in the dwarf galaxy. 
We have therefore defined a {\it young star sample} by selecting all stars in 
the {\sc cmd}\  region having $19.5 < V < 23.5$ and $B-I<0.5$.

 \begin{figure}
 \resizebox{\hsize}{!}{\includegraphics{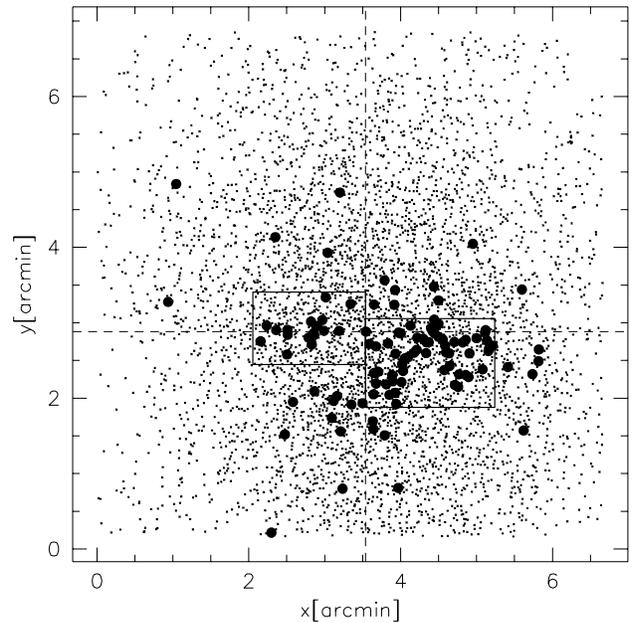}}
 \caption[]{
 Spatial distribution of the samples of red and blue stars discussed in the text
 (they are represented  by small crosses   
 and filled circles, respectively).  
 Two small boxes are used to schematically delimit the star forming region. 
 A large cross indicates the center of the galaxy. }
 \label{f_distr_spaz}
 \end{figure}

The spatial distribution of the blue star sample is plotted in 
Fig.~\ref{f_distr_spaz},  along with the distribution of stars in the red giant 
sample (defined  below).  
The galaxy center, marked by a large cross, was defined using the mode of 
the marginal distributions of star coordinates. 
The star formation sites in Phoenix are clearly delineated  in this figure. The 
young stars are found in clumps in the central region of the galaxy, the most 
prominent clump being the  well-known ``association''  described by 
Canterna \& Flower  (\cite{cant+flow77}) and studied by OG88. The spatial 
distribution of the different stellar populations in Phoenix will be further 
discussed in Sect.~\ref{s_discu}.  
This information is used here to define a ``red giant  sample'' by  excluding 
all objects in the two most prominent star forming regions (approximated by 
the two rectangles in Fig.~\ref{f_distr_spaz}).  

\begin{figure}
 \resizebox{\hsize}{!}{\includegraphics{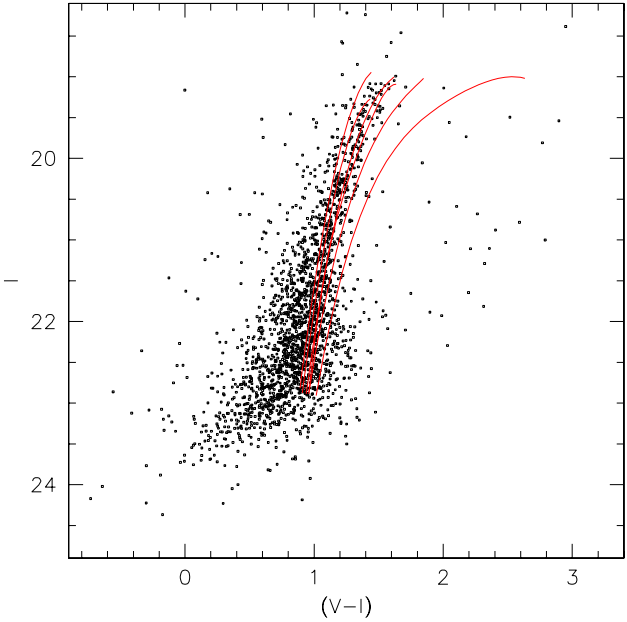}}
 \caption[]{
 The $V$, $(V-I)$\  color-magnitude diagram of stars in the RGB  sample of 
 Phoenix, which excludes all stars within the star forming regions.  Also shown are the  
 fiducial red giant branch sequences of  Galactic globular clusters from 
 Da Costa \& Armandroff (\cite{daco+arma90}), spanning a metallicity range 
 from [Fe/H]$=-2.2$\  to [Fe/H]$=-0.7$.  A distance modulus 
 $(m-M)_0=23.1$\  was adopted for Phoenix
  }
 \label{f_cmd90}
 \end{figure}

Figure~\ref{f_cmd90} presents the $I$, $(V-I)$\  color-magnitude diagram of 
the stars in the RGB sample, together with the fiducial red giant branches of  
Galactic globular clusters from Da Costa \& Armandroff 
(\cite{daco+arma90}). From blue to red, the globular clusters are M15, 
NGC~6397, M2, NGC~6752, NGC~1851, and 
47~Tuc, whose metallicities  are  
${\rm[Fe/H]}=$--2.17, --1.91, --1.58, --1.54, --1.29, and --0.71 dex, 
respectively. We note the extremely well defined RGB cutoff of Phoenix, 
suggesting the lack of a  dominant  intermediate age population.  

 \begin{figure}
 %
 \resizebox{\hsize}{!}{\includegraphics{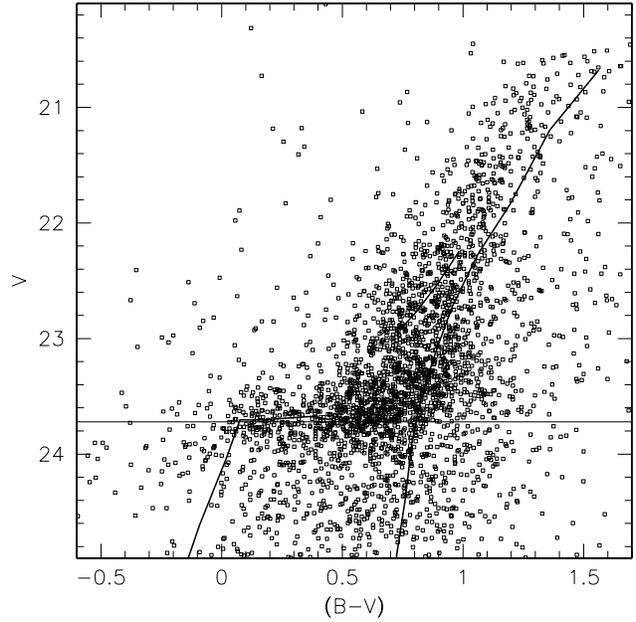}}
 \hspace{0cm} %
 \caption[]{
 The color-magnitude diagram of Phoenix stars  farther 
 than $1\farcm3$\ from the galaxy center. 
 A horizontal branch extended to the blue is evident at $V=23.73$\ 
 (this level is indicated by a horizontal line). 
 Also shown are the fiducial sequences  of RGB, HB, and AGB stars in the 
 {\sc cmd}\   of  the Galactic globular cluster  M3 (from Buonanno et al. 
 \cite{buon+88})}
 \label{f_hb}
 \end{figure}

An expanded view of the horizontal branch of Phoenix is provided in 
Fig.~\ref{f_hb}.  This diagram was obtained by picking only stars in 
the outer region of the galaxy ($R>1\farcm3$), thus excluding all 
objects in the star formation regions.  
This selection proved to enhance the contrast between the HB and the RGB, 
partly because photometric error are smaller in the outer, less crowded 
regions, but possibly also because of an intrinsic difference in the radial 
distributions of  the RGB and HB populations (we will return to this point in 
Sect.~\ref{s_discu}).  Figure~\ref{f_hb} clearly confirms the presence of a 
moderately blue horizontal branch in Phoenix at $V\sim23.7$. We note that 
the HB is hardly detected  in the $V$, $(V-I)$\ diagram because most of the 
faint, blue HB stars fall below the detection threshold in the $I$\ band.  
The fiducial loci of  the Galactic globular cluster M3 (which appears 
obviously more metal-rich than Phoenix: 
[Fe/H]$=-1.57$) are shown for comparison purposes, appropriately shifted to 
the Phoenix distance as  derived from the RGB cutoff   (see 
Sect.~\ref{s_dista}). The data for M3 are from Buonanno et al. 
(\cite{buon+88}), and the apparent distance modulus and reddening from 
Harris (\cite{whar98}). 

\section{Distance and metallicity} 
\label{s_analy}

The sample of red giant branch stars used to construct the Phoenix RGB 
luminosity function  (LF)   is a subset of  the master red sample defined in 
the previous section. To reduce the contamination by young stars and field 
objects,  we further selected  
a  ``2$\sigma$\  RGB sample'' which comprises all stars within 
$\pm 2\sigma$ from the fiducial ridge line. The details of the method are 
given in Sect.~\ref{s_rgbwidth}. 
The stars in the $2\sigma$ red giant sample are used below to derive 
the fundamental parameters of Phoenix, distance and metal abundance.

\subsection{Luminosity function and distance based on the RGB tip}
\label{s_dista}

The distance to Phoenix was first estimated from the $I$ absolute magnitude 
of the tip of the red giant branch (see Madore \& Freedman 
\cite{mado+free95} for a discussion of the method and previous work). 
The cutoff was found at 
$I_{\rm tip} = 19.09 \pm 0.05$.  The  error mainly reflects the uncertainty in 
locating  the RGB tip, which is larger than the systematic error on our $I$\ 
magnitude scale. 
To assess  the  effects of crowding on the estimated  RGB cutoff, 
we simulated the upper part of the RGB using artificial stars in the 
{\sc cmd}\  region $1.3<V-I<1.7$,  $I \geq 19.0$.  After processing the artificial 
data as the real RGB stars, the cutoff magnitude of the retrieved stars 
was 0.03 mag too bright, independently of  the   radial distance from the 
galaxy center.  A correction for this small bias was included in our 
estimation of  the distance modulus. 

The bolometric and $I$\ luminosities of the RGB tip, 
$M_{\rm bol}^{\rm TRGB}$\ and  $M_I^{\rm TRGB}$, and the $I$\ 
bolometric correction, were then  estimated using  the relations of Da Costa 
\& Armandroff (\cite{daco+arma90}).
From the extinction-corrected color of red giants near the tip, 
$(V-I)_{\rm 0,tip} = 1.48 \pm 0.04$, we obtain $BC_I = 0.52 \pm 0.01$. 
The bolometric absolute magnitude of the tip was then inferred from our 
estimate of the mean metal abundance of Phoenix, 
${\rm [Fe/H]} = -1.81\pm 0.10$  (cf. Sect.~\ref{s_metal}).
The result was  $M_{\rm bol}^{\rm TRGB} = -3.47 \pm 0.02$ (the error 
reflects the uncertainty on metal abundance), from which 
$M_I^{\rm TRGB} = -3.99 \pm 0.02$\  was finally obtained. 
For  the extinction and reddening  corrections we used 
$A_V =3.1 \, E_{B-V} = 0.06 \pm 0.06$\  and 
$E_{V-I} =1.28 \, E_{B-V}=0.03\pm0.03$, 
adopting $E_{B-V} = 0.02$\   mag  from Burstein \& Heiles         
(\cite{burs+heil82}), with a $\pm0.02$\  mag reddening uncertainty. 
The distance modulus to Phoenix  determined from the 
extinction-corrected magnitude of the  RGB tip is then 
$(m-M)_0 = 23.04 \pm 0.07$, where the error includes the photometric and 
reddening uncertainties, and the uncertainty on $M_I^{\rm TRGB}$.  
This distance modulus is slightly shorter than that derived by VDK91 using 
the same method and reddening. 


\subsection {Distance based on horizontal branch stars }

 \begin{figure}
 \resizebox{\hsize}{!}{\includegraphics{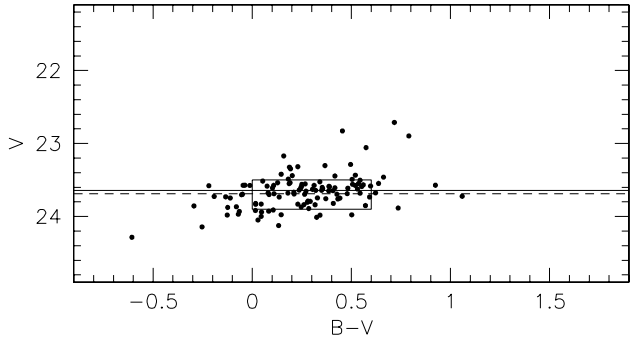}}
 \caption[]{
 A simulation of the effects of crowding and photometric errors for  stars 
 with magnitudes and colors typical of  the observed HB. Artificial stars 
 originally distributed inside the outlined box are spread over a larger color 
 range after processing. The median of the input and output $V$\ magnitudes 
 are represented by the dashed and solid line, respectively 
 }
 \label{f_hbsimu}
 \end{figure}

Our first detection of the HB of Phoenix allows to obtain an independent 
estimate of the distance to Phoenix. 
The mean  magnitude of  the horizontal branch was calculated as the  
median $V$\ magnitude of the 553 HB stars in the {\sc cmd}\  region  
$0.0 < B-V < 0.6$,  $23.4 < V < 24.0$. 
The median is  $V_{\rm HB}=23.73$, with a r.m.s. scatter 
0.17  mag,  yielding a formal error on the mean  of ${\la}0.01$ mag. 
The mean value $V_{\rm HB}$\  was found to remain constant for different 
radial sub-samples, within the internal errors.  The total uncertainty, 
including the systematic error of the $V$\ magnitude scale,  is $\sigma_V = 0.016$\  
mag. 

Given the significant incompleteness of our photometry at the HB level, the 
effects of errors and incompleteness on the measured location of the HB need 
careful  investigation. The results of our artificial star experiments were 
used to simulate the HB as a narrow strip in the {\sc cmd}\  
($23.5<V<23.9$, $0.0<B-V<0.6$).   The distribution of the retrieved artificial 
stars in the color-magnitude diagram (Fig.~\ref{f_hbsimu}) 
was then analyzed exactly in the same way as the real HB data. The median  
HB level of the retrieved  stars was $V=23.65$, {i.e. }
the simulated HB appears slightly biased toward brighter magnitudes due to 
the rapidly changing photometric completeness, though the effect is 
relatively modest.  Taking into account this bias, we adopt
$V_{\rm HB}=23.78 \pm 0.05$\  as the mean observed magnitude of the blue 
HB. 

Using this value for $V_{\rm HB}$, we calculated the distance modulus of 
Phoenix on the Lee {{et al.}\ } (\cite{ywlee+90}) distance scale,  using their relation 
for the absolute visual magnitude of RR Lyrae variables, 
\begin{equation}
M^{\rm RR}_V=0.17~ {\rm [Fe/H]} + 0.82
\end{equation}

for a helium abundance of $Y=0.23$.  We adopt this calibration since it is 
also the basis of the RGB tip method (Da Costa \& Armandroff 
\cite{daco+arma90}; Lee {{et al.}\ } \cite{mglee+93}).   
On the  Lee's  {{et al.}\ }  scale,  which gives
$M^{\rm HB}_V =M^{\rm RR}_V=0.51$\  mag 
for a metallicity [Fe/H]$=-1.81\pm 0.10$\  ({{cf.}\ } Sect.~\ref{s_metal}),
we estimate a distance modulus $(m-M)_0=23.21 \pm 0.08$. 
Note that the distance error includes the photometric (statistical and 
zero-point) errors and the reddening uncertainty, but does not take into 
account the uncertainty of the adopted HB calibration and distance scale. 
For example, 
using the empirical calibration of the mean absolute  magnitude of the HB in 
8 M31 globular clusters by Fusi-Pecci {{et al.}\ } (\cite{fusi+96}), the distance 
modulus would be $(m-M)_0=23.01 \pm 0.10$, {i.e. } brighter by about 0.2 mag. 

The distance based on the HB is only marginally larger than that determined 
from the tip of the RGB, and the agreement is good in view of the many 
sources of uncertainty.  The average of the two determinations, 
$(m-M)_0=23.12\pm 0.08$,  corresponding to $421 \pm 16$\ kpc, 
is finally adopted as our best estimate of the distance to Phoenix.

\subsection{Mean abundance}
\label{s_metal}

 \begin{figure}
 \resizebox{\hsize}{!}{\includegraphics{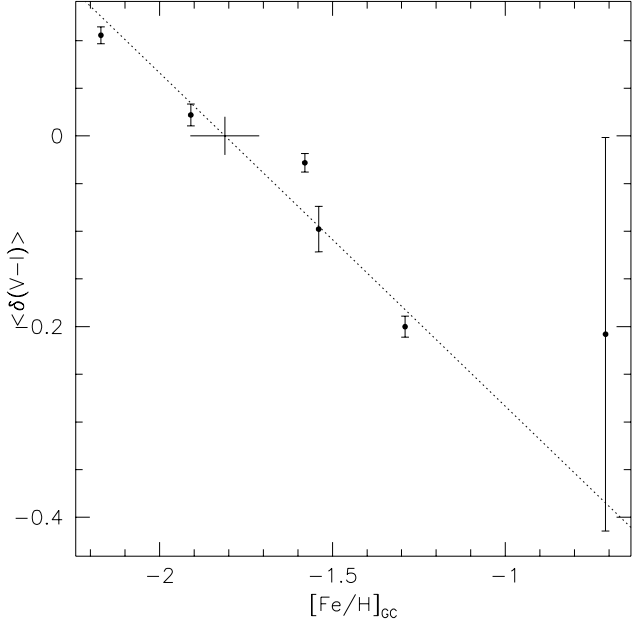}}
 \caption[]{
 The mean color difference of the Phoenix red giant stars from the fiducial 
 sequences of template globular clusters  is plotted  against the mean cluster 
 metallicity.   The error bars represent  the standard deviations of the 
 residuals in  $(V-I)$. Also shown is a linear regression in the range  $-2.2 
 <{\rm [Fe/H]} < -1.3$\  (dotted line). 
 The large cross identifies  the mean abundance of Phoenix together with its 
 error. 
 }
 \label{f_fehdvi}
 \end{figure}

As in Paper~I, the mean metal abundance of Phoenix  was estimated by 
direct comparison of the red giant branch in the $V$, $(V-I)$\ 
color-magnitude diagram with the RGB ridge lines of the template globular 
clusters from Da Costa \& Armandroff (\cite{daco+arma90}).  We calculated 
the mean color differences $\delta(V-I)_0$ between the data in the Phoenix 
red giant sample and the fiducial loci of the template clusters, using different 
choices for the luminosity range.  The values of $\delta(V-I)_0$\ were 
corrected for 
extinction  adopting  $E_{V-I} = 0.03 \pm0.03$. 
We note that for magnitudes fainter than $I\sim20.5$, the color distribution 
of the Phoenix RGB is skewed towards bluer colors,  due to the presence of 
the unresolved  AGB of the  old population (a similar trend was noticed  in  
the {\sc cmd}\  of And~I  by Da Costa  {{et al.}\ }  \cite{daco+96}).
In Galactic globular clusters, AGB stars can be separated from 
red giant branch stars below an absolute magnitude 
$M_V \approx -1.5$. 
In Phoenix this precisely corresponds to $V \approx 21.5$\ or 
$I\approx20.5$. 
 
Figure~\ref{f_fehdvi} shows the mean color residuals $\delta(V-I)_0$ 
(calculated in the range $-4 <M_I <-3$)  against the metallicities of the 
Galactic clusters. A linear fit to the  differences, excluding only  the metal-
rich cluster 47\,/Tuc, gives   the relation

\smallskip\noindent
\begin{equation}
{\rm [Fe/H]} = -2.86~\delta(V-I)_0   -  1.81, 
\label{e_vimetal}
\end{equation}

\smallskip\noindent
applicable in the range $-2.2<{\rm [Fe/H]}<-1.3$.  
This relation implies that the interpolated metal abundance  of Phoenix is 
[Fe/H] $= -1.81\pm 0.10$ dex.  The abundance ($1\sigma$) error is derived 
from the total uncertainty on the mean $(V-I)$\ color using 
Eq.~\ref{e_vimetal}. 
The uncertainty on $\delta(V-I)_0$\ includes the statistical error  ($\sim 
0.01$\  mag),  the systematic error  due to calibration uncertainties (0.02  
mag), 
and the reddening uncertainty. 
We do not include  the uncertainties related   to  systematic differences between  
abundance scales for the Galactic globular clusters.  Note that for metal-poor  
populations, however, there is reasonable agreement between  different  abundance 
scales ({{e.g.},\ }  Carretta \& Gratton \cite{carr+grat97}). 

The mean abundance derived here is slightly higher than the value measured 
by van de Rydt {{et al.}\ } (\cite{vdry+91}). The difference is probably to be 
ascribed to the small shift  in the calibrated $(V-I)$\  colors ({{cf.}\ } 
Sect.~\ref{s_obsred}). 
Using  a wider magnitude range for calculation of the 
$\delta(V-I)_0$\ values, a slightly lower metal abundance would have been 
derived ({{e.g.},\ }  [Fe/H]$=-1.92$\ using the upper 2 mag of the RGB). This 
reflects the bias in the giant branch color   due to the presence of AGB stars. 
The difference is comprised in the quoted uncertainties, though. 

\subsection{Metallicity dispersion}
\label{s_rgbwidth}

 %
 \begin{table}[h]
 \caption{Mean $V-I$\  and color dispersion along the RGB 
 \label{t_sigmas}}
\begin{flushleft}\begin{tabular}{lrrrr}
 \noalign{\smallskip}\hline\noalign{\smallskip}
 \multicolumn{1}{c}{$I$} &
 \multicolumn{1}{c}{$V-I$} &
 \multicolumn{1}{c}{$\sigma_{V-I}$} &                      
 \multicolumn{1}{c} {$\sigma_{V-I}$} &
 \multicolumn{1}{c} {$\sigma_{V-I}$} \\
   & & RGB &  simul. & intrinsic \\
 \noalign{\smallskip}\hline\noalign{\smallskip}
  19.25 & 1.43 &  0.090  &  0.044  & 0.079 \\ 
  19.75 & 1.33 &  0.079  &  0.049  & 0.093 \\ 
  20.25 & 1.22 &  0.079  &  0.052  & 0.059 \\ 
  20.75 & 1.11 &  0.100  &  0.052  & 0.085 \\ 
  21.25 & 1.03 &  0.150  &  0.072  & 0.137 \\ 
  21.75 & 0.95 &  0.153  &  0.087  & 0.126 \\ 
  22.25 & 0.90 &  0.183  &  0.127  & 0.132 \\ 
  22.75 & 0.80 &  0.217  &  0.163  & 0.143 \\ 
 \noalign{\smallskip}\hline\end{tabular}\end{flushleft}\end{table}

 \begin{figure}
 \resizebox{\hsize}{!}{\includegraphics{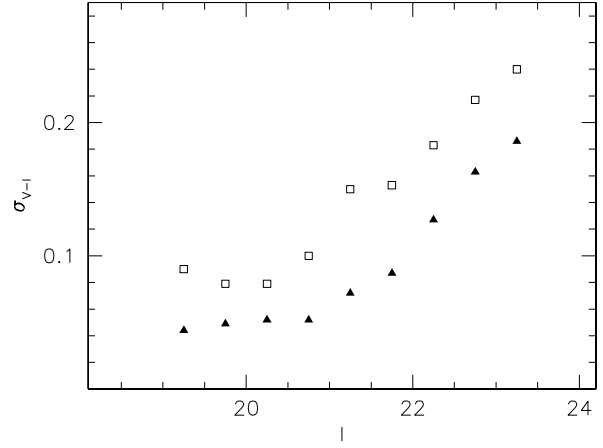}}
 \caption[]{The $(V-I)$\  color dispersion of red giant stars in Phoenix 
 in 0.5 mag bins (open squares), plotted together with the 
 instrumental color scatter derived from simulated stars (filled triangles) 
 }
 \label{f_visigmas}
 \end{figure}

Previous work has suggested  the presence of an intrinsic color width in the 
red giant branch of Phoenix.  We find confirmation of this claim in our data. 
Table~\ref{t_sigmas} gives our results for the mean color and measured 
RGB width at different luminosities (columns 2 and 3). The mode and 
standard deviation of the $(V-I)$\  color distribution of red giants were 
measured in 0.5  $I$-mag  bins on a rectified version of the RGB, using a 
Gaussian fit with a $3\sigma$\  clipping to discard field objects and young 
stars.  
The instrumental scatter derived from artificial star experiments is given in 
column 4.   The color dispersion of simulated stars, 
distributed along the RGB fiducial locus, was measured with the same 
robust approach used to derive the RGB widths.  
The observed color dispersions and instrumental errors 
are compared in Fig.~\ref{f_visigmas}. 
Clearly, the  RGB  color scatter is  significantly larger  at any luminosity  
than expected  from measurement errors alone. 
 The intrinsic $(V-I)$\ dispersions calculated as the quadratic difference 
between the measured and instrumental scatter are given in col.5 of  
Tab.~\ref{t_sigmas}.   The average  color dispersion in the 4 brightest 
magnitude bins ($I<21.0$) is $0.079\pm0.008$\ (r.m.s. error on  the mean). 
It is interesting to note that the values are roughly constant above $I=21$. 
Below  this magnitude, the presence of AGB stars evolving away from the 
horizontal branch precludes reliable analysis. 

This evidence for an intrinsic scatter has been independently confirmed by 
a direct comparison of our photometry with the data of van 
de Rydt {{et al.}\ } (\cite{vdry+91}).  In the range
$19.0 <  I  <  19.5$, the color differences for the stars in common  yield 
a standard deviation of 0.072 mag. 
Under the simple assumption that the two samples have comparable errors, 
we obtain an instrumental scatter of  0.051 mag, which implies  0.074 mag 
for the intrinsic scatter, in  good agreement with that obtained from 
crowding simulations. We thus conclude that an intrinsic color scatter is 
indeed present in Phoenix, of the order 
$\sigma_{\rm intr}(V-I)=0.08\pm0.01$\  mag. 
If we are to explain this scatter as due to an abundance spread, it would 
correspond to  a metallicity dispersion 
$\sigma_{\rm [Fe/H]}=0.23 \pm 0.03$\  dex (using Eq.~\ref{e_vimetal}).  
Our data thus seem to indicate that 
the  RGB stars in Phoenix span a ($\pm1\sigma$) range in metal abundance  
 $-2.04 < {\rm [Fe/H]}  <  -1.58$\   dex, {i.e. } a range of ${\la} 0.5$\  dex.
This confirms the relatively modest abundance range found by OG88, 
while VDK91 give a somewhat larger metallicity spread.

 An abundance scatter similar to that found in the Phoenix dwarf is well 
established in many dwarf spheroidals, where it has been confirmed also by 
spectroscopic observations (Suntzeff et al. \cite{sunt+93}; Da Costa 
\cite{daco98}).  In dSph's, a range in metal abundance is indicative of an 
enrichment process  in which multiple stellar generations have taken place 
from interstellar gas enriched by previous episodes.
However, a range in age may also affect the RGB color dispersion. 
As an example, we have calculated the mean $(V-I)$\  color difference 
between  two isochrones with ages of 15 and  5 Gyr respectively,  
using the models of  Bertelli {{et al.}\ } (\cite{bert+94}).
The shift was calculated in  the 
luminosity range   $-4 < M_I <  -3.5$, {i.e. } the 
interval used for calculating $\sigma_{\rm [Fe/H]}$, for two different 
metallicities,  $Z=0.0004$\  ([Fe/H]$=-1.7$) and  $Z=0.004$\ ([Fe/H]$=-0.7$). 
  We find a mean shift 
$\Delta(V-I)\approx0.07$\  for the metal-poor isochrones and 
$\Delta(V-I)\approx0.03$\ for the metal-rich ones (younger stars are bluer). 
The effect of an age mix on the $(V-I)$\ colors of the red giants is therefore 
not negligible, particularly  in the case of  metal-poor isochrones, for which 
the color difference near the RGB tip mimics a metallicity variation of  $\sim 
0.2$\  dex. 
Thus, depending on the details of the star formation and enrichment history 
of each galaxy, the effects of  a younger age and  a higher metallicity  may 
partly cancel out, in which case the abundance spread  inferred from the 
color dispersion would  {\it underestimate} the metallicity  range. An 
extreme example of such a combined  effect  may be at work in the Carina 
dwarf, 
which has a very narrow RGB  indicative of  little chemical enrichment (of 
the order 0.2 dex) although it has had several generations of stars 
(Smecker-Hane {{et al.}\ } \cite{smec+94}; Hurley-Keller {{et al.}\ } \cite{hurl+98}).  
Understanding the observed color-magnitude diagrams of a complex stellar 
system requires realistic modeling of both its star formation and chemical 
enrichment histories (best performed on high-resolution data), and 
independent information on metal abundances such as that provided by 
spectroscopy of individual stars. 

\section{The stellar content of Phoenix}
\label{s_discu}

\subsection{The old population: horizontal branch morphology}

We begin our analysis and discussion of the star content of Phoenix with a 
few comments on our HB detection.  The moderately blue horizontal branch 
unquestionably implies that Phoenix harbors a sizable old stellar population.  
Due to the relatively large photometric errors,  stars on the red HB cannot 
be disentangled from the RGB, yet we can obtain  some information 
about  the HB morphology looking at Fig.~\ref{f_hb}. 
The horizontal branch of Phoenix appears mostly populated on the red side 
and  moderately  extended to the blue ($0.0<B-V<0.8$), with a hint of the RR 
Lyrae gap  at  $B-V\sim0.4$.  
Thus it  appears  remarkably similar to the HB types  of  the dwarf 
spheroidals  Leo~II 
(Demers \& Irwin \cite{deme+irwi93};  Mighell \& Rich \cite{migh+rich96}),    
And~I (Da Costa {{et al.}\ } \cite{daco+96}),
Draco  (Grillmair {{et al.}\ } \cite{gril+97}),  and Tucana (Seitzer {{et al.}\ } 
\cite{seit+98}; see also Da Costa \cite{daco98}).   
Given the mean metallicity of Phoenix, this HB morphology implies a mild 
second parameter effect, since old halo clusters 
with [Fe/H]$\sim-1.8$\  have blue horizontal branches.  
Under the hypothesis that the HB morphology of Phoenix is mainly 
driven by age, the theoretical HB models of  Lee {{et al.}\ } (\cite{ywlee+94}), and 
in particular their HB-type versus metallicity  diagram,  would indicate for 
its old population an age  $\sim2-3$\ Gyr younger than that of  the old halo 
clusters. However, it appears most likely that the HB of Phoenix is due to 
a mixture of stars of different ages.

Recent studies have shown  that the blue HB population in some dSph's 
({{e.g.},\ } And~I and  Fornax)  is less centrally concentrated than the red HB stars 
and the red giants, most likely due to a gradient in the mean age of the 
galaxy populations 
({{e.g.},\ } Da Costa {{et al.}\ }  \cite{daco+96}; Stetson {{et al.}\ } \cite{stet+98}).  
To investigate the presence of a radial population gradient in Phoenix, 
we have compared the radial distribution of the blue horizontal branch stars 
($23.4< V <24.0$, $0.0<B-V<0.6$) to that of red giants just above the HB 
($22.8< V <23.4$, $0.5<B-V<1.1$).  
The surface density profile of the HB stars appears to be  
substantially  more extended  than that of red giants.   
A two-sided Kolmogorov-Smirnov test clearly indicates that the  two spatial 
distributions differ at a 99\% confidence level. 
Since there might be some concern that apparent gradients  be induced  
by different  crowding and completeness in the two magnitude and color 
ranges, we applied the same test to the radial distributions of 
simulated stars in the same {\sc cmd}\  regions. 
The cumulative radial density profiles of the blue-HB and RGB artificial
stars look very similar. Formally, the null hypothesis (the two data sets  
are drawn from the same parent population) cannot be rejected 
at any significant confidence level.
We therefore  conclude that the detected  gradient cannot be produced by 
instrumental effects. 

This extended distribution of the blue-HB sample  ({i.e. } of  the  oldest stars) 
can have different implications. 
A radial change in the horizontal branch morphology could be due either to 
a radial change in the mean age of the stellar populations, or to a higher 
mean abundance in the central regions, or both.  Clearly, it is not easy to 
decide between these alternatives on the basis of our data (the radial 
dependence of the HB morphology will be better investigated  using HST 
photometry). We just note that the observed central concentration of  young 
and intermediate age stars (see Sect.~\ref{s_young}) indicates that multiple 
star formation episodes occurred preferentially in the central regions. 

\subsection {The intermediate age population: AGB stars}
\label{s_discu_agb}

 \begin{figure}
 \hbox{
 \resizebox{\hsize}{!}{\includegraphics{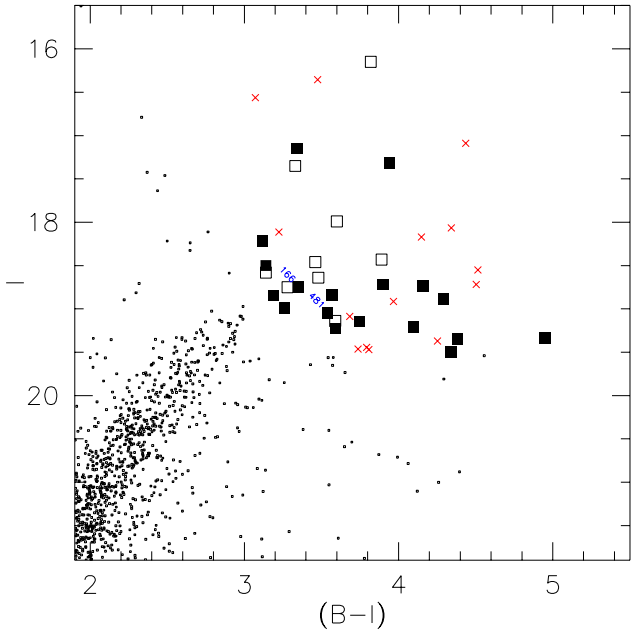}}
 }
 \caption[]{AGB star candidates in Phoenix, selected as redder than 
 $(B-I)=3.0$\ and brighter than $I=19.5$\ (squares).  
 Open symbols identify possibly blended stars. The two carbon stars 
spectroscopically identified by Da Costa (\cite{daco94}) are indicated 
by their numbers in VDK91.
 {\em Crosses} represent  field contamination  in the same region of the {\sc cmd}
 }
 \label{f_agbplot}
 \end{figure}

Previous studies of Phoenix have found a few very red stars located above 
the red giant branch tip. These may be AGB stars belonging to an
intermediate age component. 
Da Costa (\cite{daco94}) obtained spectroscopic confirmation of two carbon 
stars with $M_{\rm bol}\approx -3.7$\  in Phoenix.  These two confirmed C 
stars are less luminous than the brightest carbon stars in the  $\sim8$\  Gyr 
old SMC cluster Kron~3, and considerably fainter than those in the 
populous young (1--3 Gyr old) clusters in the LMC.  This would indicate that 
Phoenix did not form many stars since 8--10 Gyr ago  (Da Costa 
\cite{daco98}). 
This suggestion, however,  is based on a very limited sample of C stars, and 
brighter intermediate age stars may be identified  with a complete census 
of possible upper-AGB stars in this galaxy. 

The relatively large field  investigated in this paper gives us the possibility to 
better assess the contribution of an intermediate age population in Phoenix, 
and to obtain some constraints on the  star formation history of this dwarf 
galaxy. 
To this purpose, we selected stars brighter and redder than the RGB tip 
($I<19.5$,  $B-I>3.0$) as candidate upper-AGB stars younger of 
$\sim10$\ Gyr  (Fig.~\ref{f_agbplot}). 
The lower  magnitude limit,  $V=19.5$, was chosen slightly fainter of the 
RGB tip to account for 
possible large magnitude errors due to blending.  All selected 
stars were visually inspected, and those with elongated shapes or other hints 
of  nearby companions were flagged as less reliable, although further study 
(in particular spectroscopy)  will be needed to confirm  the nature of  any  
candidate. 
Figure~\ref{f_agbplot} also shows the  foreground and background objects 
found in the same color and magnitude range (crosses).  The brightest  stars 
are  obviously in  excess over field objects. 

Taking into account field contamination and incompleteness, we counted 11 
AGB  stars  in the {\sc cmd}\  region just above the RGB tip ($18<V<19$), in the 
color range  $3.0<(B-I)<4.4$.  This interval is equivalent  to a range in 
bolometric luminosity between $M_{\rm bol}=-3.5$\ (the RGB tip, where 
AGB evolution terminates in metal-poor globular clusters) and  
$M_{\rm bol}=-4.5$, this brighter end roughly corresponding  to 
the maximum AGB luminosity reached  by 3 Gyr old stars 
(Frogel et al. \cite{frog+90};  Marigo et al. \cite{mari+96}). 
For comparison, 156 RGB$+$AGB stars are 
counted within 1 mag below the RGB tip ($19<I<20$).  

The occurrence of  upper-AGB stars in dSph's generally indicates the 
presence of a population significantly younger than that of galactic globular 
clusters, and this seems to be the case also for Phoenix. 
Alternative explanations for the stars brighter than the RGB tip seem 
unlikely. For instance, large amplitude long-period variable stars (LPV's) 
have not been found in old halo globular clusters as metal-poor as Phoenix 
(Frogel \& Elias \cite{frog+elia88};  Frogel \& Whitelock \cite{frog+whit98}).  
Even accounting for the modest abundance dispersion inferred in 
Sect.~\ref{s_rgbwidth}, there is no evidence for a significant component in 
Phoenix  with metallicity higher than [Fe/H]$\approx-1$.  
We have also considered the possibility that some of our AGB candidates are 
artifacts of blending of red giants near the tip (Renzini \cite{renz98}). 
 Assuming  that the luminosity of Phoenix ($M_V\approx-9.7$, van de Rydt 
{{et al.}\ } \cite{vdry+91}) is uniformly distributed over a $4\arcmin \times 
4\arcmin$\  area, and adopting a resolution element 1.7  $\sq\arcsec$,  
we expect less than one resolution element (0.4) in our frame 
to contain two red giants near the RGB tip.
Further, in our simulations we find just 1 star scattered above the 
RGB tip by photometric errors. 

Thus we conclude that most of the stars brighter than $I=19$\ trace the 
extended AGB of an intermediate age population. Our counts of upper-AGB 
stars are used to  estimate the contribution of  the intermediate age 
component  to the luminosity of Phoenix, following the methods of Renzini 
\& Buzzoni (\cite{renz+buzz86}) and Renzini (\cite{renz98}). 
The ``fuel consumption theorem''  gives 
the number of stars in a post-MS evolutionary phase $j$, 
$n_j =B(t)~L_T~t_j$, 
as a function of the total bolometric luminosity of the population ($L_T$) 
and the lifetime of the phase ($t_j$). The {\it specific evolutionary flux} 
$B(t)$   is of the order $2\times10^{-11}$ stars  {$L_{\odot}$}$^{-1}$\  yr$^{-1}$\  in 
good approximation for any age between 3 and 10 Gyr. Even though 
evolutionary predictions in Renzini (\cite{renz98}) are given for a 
stellar population with solar composition, the results for $n_j $\  are almost 
unchanged for metal-poor systems because a lower metallicity has opposite 
effects on the bolometric corrections and lifetimes of red giants 
at a given age (Maraston 1998, priv. comm.). 
For thermally pulsing AGB stars we assume a standard lifetime of  1 Myr for each 
magnitude of  luminosity increase (Renzini \cite{renz98}). 
An absolute magnitude $M_V\simeq-9.7$\ is adopted  for  the red populations of 
Phoenix,  corresponding to a total luminosity  $L_V = 6.2\times10^5$\ {$L_{\odot}$}\ 
(van de Rydt {{et al.}\ }  \cite{vdry+91}). 
If  the bulk of the Phoenix population were of intermediate age ($3< t <10$\ 
Gyr), then the expected ratio between  the number of thermally pulsing AGB 
stars in the 1 mag interval above the tip, and  the sum of RGB and 
early-AGB stars in the 1 mag range below the tip, would be 
$N_{\rm AGB}$/$N_{\rm RGBT} \approx 0.19$.  
This is about 3 times the fraction observed in Phoenix,  where 
$N_{\rm AGB}$/$N_{\rm RGBT}=11/156=0.07\pm 0.03$\ 
(the error simply reflects the count statistics). 
The number of  upper-AGB stars  is therefore consistent with  
$\sim37\pm12$\% of the stellar population in Phoenix being 
of intermediate age, {i.e. }  even the  small observed number of AGB stars 
implies a significant  intermediate age population. 
This not surprising since the lifetime of the 
thermally pulsing AGB phase is short enough that a large  sample of 
progenitors is required to observe an appreciable number of 
upper-AGB stars.  
The possible concentration of stars on the 
red side of the HB is indeed consistent with the presence of an 
$\sim8$\ Gyr component, although a firm quantitative estimate 
of the fraction of intermediate age stars, 
and their mean age, must await for deeper images.

\subsection {The young stars}
\label{s_young}

 \begin{figure}
 %
 \resizebox{\hsize}{!}{\includegraphics{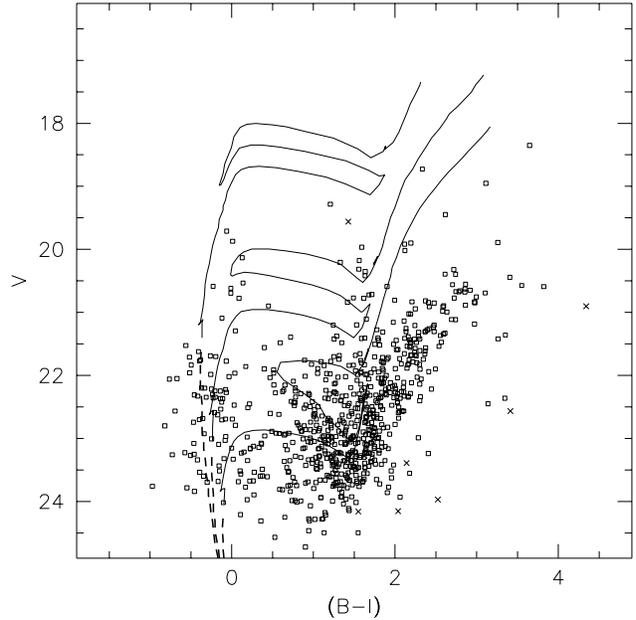}}
 \caption[]{The color-magnitude of Phoenix in the star-forming regions, 
 compared to a set of theoretical isochrones from Bertelli et al. 
 (\cite{bert+94}), with metallicity $Z=0.0004$\ and ages (1.0, 2.5, 6.3) 
 $\times10^8$\ yr. 
 The {\em dashed lines} represent the main sequence evolution.  
 Also shown are the foreground/background objects in an equal area of the 
 comparison field  ({\em crosses})
 }
 \label{f_sf_reg}
 \end{figure}

We now turn to one of the most distinctive features of Phoenix, the presence 
of a recent burst of star formation. The wide baseline of the $(B-I)$\ color 
used in  this study, and the availability of a comparison field,  allow us a 
better  separation of stars in different evolutionary phases. 
Figure~\ref{f_sf_reg} presents our  {\sc cmd}\  for stars comprised within the star 
formation regions (the rectangles in Fig.~\ref{f_distr_spaz}). 
Superposed on the diagram are three representative  isochrones from the 
Padua stellar evolution models (Bertelli {{et al.}\ } \cite{bert+94}),  for a metal 
abundance $Z=0.0004$\ ([Fe/H]$\approx-1.7$) 
and ages  $1.0\times10^8$, $2.5\times10^8$\ yr, 
and $6.3\times10^8$\ yr.  The corresponding masses at the MS 
turnoff  are 4.8, 3.0, and 2.0 {$M_{\odot}$}, respectively.   
The model colors are 
sensitive to the adopted metallicity particularly in their evolved part. 
A  $\sim10^8$\  yr old model provides a good fit to the 
upper main sequence of the young population in Phoenix. 
While in general core helium burning (HeB) stars on the blue loops give 
a substantial contribution to the blue plume in ground-based photometry
({{e.g.},\ } Tosi {{et al.}\ } \cite{tosi+91}), the gap 
at $(B-I)\approx0.5$\ in Fig.~\ref{f_sf_reg} 
suggests that we have reached a 
good  separation of the hydrogen and HeB star sequences. 
The group of bright blue stars having $(B-I)\sim0.5$\ and  $V<21.5$ 
 matches quite well the expected location of HeB stars belonging 
to a 250  Myr old population.  
This suggestion is confirmed by the presence of a distinct 
sequence at $(B-I)\sim1.5$\  which  appear to trace the red 
end of the blue loops.  
Also,  a short  sequence of stars comprised in the region 
$0.5\leq (B-I) \leq1.5$,  $22\leq V \leq23$\ is  
quite  well matched by the blue loop phase of the  $6\times10^8$\  yr old 
isochrone.  The brightest red stars ($V \leq20$)  almost overlap the red 
supergiant phase of the same  $6\times10^8$\  yr isochrone 
(although the precise location of these luminous stars is a sensitive function of 
metallicity).
Thus, the emerging picture is one in which the most recent star formation 
episode in Phoenix had finite duration (of the order 0.5 Gyr or more), with 
evidence of a distinct burst  $\sim$0.6 Gyr ago, and went on forming stars 
until  about $10^8$\ yr ago. It is interesting to note that this picture is similar 
to  what we have found  in Fornax,  where star formation proceeded from 1 
Gyr until 0.2 Gyr ago (Held {{et al.}\ } \cite{held+99}).

How significant is this  young population in terms of luminosity and mass ?  
To estimate the $V$\  luminosity of the young star component, we selected 
the blue stars  in the color and magnitude range  $-1<B-I<0.5$,  
$21.5<V<23.0$\ 
(the faint limit roughly corresponds to the MS turnoff of the 0.25 
Gyr isochrone). 
The stars were counted in 0.5 mag bins, and corrected for incompleteness 
according to the results of  the artificial star experiments.
We  found 57 blue stars, corresponding  to 106 stars  after completeness 
correction. 
Assuming that all these are MS stars distributed according to a 
Salpeter  initial mass function, an order-of-magnitude estimate of the $V$\ 
luminosity of the youngest stars was obtained following Renzini 
(\cite{renz98}), under the simple assumption of a single 100 Myr old burst of star 
formation. The observed stars are predicted in the mass  range  
$3.0 <M< 4.8$\  {$M_{\odot}$}\   for a total $V$\  luminosity of  
$L_{V} \approx 3.6\times10^4$\  {$L_{\odot}$}. 
Using the bolometric corrections and  mass-to-light ratios of  Maraston 
(\cite{mara98}),  these correspond  to $\sim 5.6 \times10^4$\ {$L_{\odot}$}\  in 
bolometric units or a mass of  $\sim7\times 10^3$\ {$M_{\odot}$}\ for the most recent burst.  
The youngest stars thus contribute  ${\la}$6\% to the $V$\ luminosity of 
the galaxy, corresponding to a  mass fraction 
$M_{\rm young}/M_{\rm old}\approx 0.002$.
These results are similar to those 
obtained by Mould  (\cite{moul97}) and Aparicio et 
al. (\cite{apar+97})  for LGS3, a dwarf galaxy which shares the basic 
characteristics of Phoenix, {i.e. } a smooth optical appearance 
accompanied by  recent star formation.  
These estimates do not include the blue stars older than $250$\ Myr, 
in particular the 600 million yr component.  
The star formation history of Phoenix (in particular in the last 1 Gyr)  
will be studied in more detail in a future paper 
using deeper data and synthetic {\sc cmd}'s

 \begin{figure}
 \resizebox{\hsize}{!}{\includegraphics{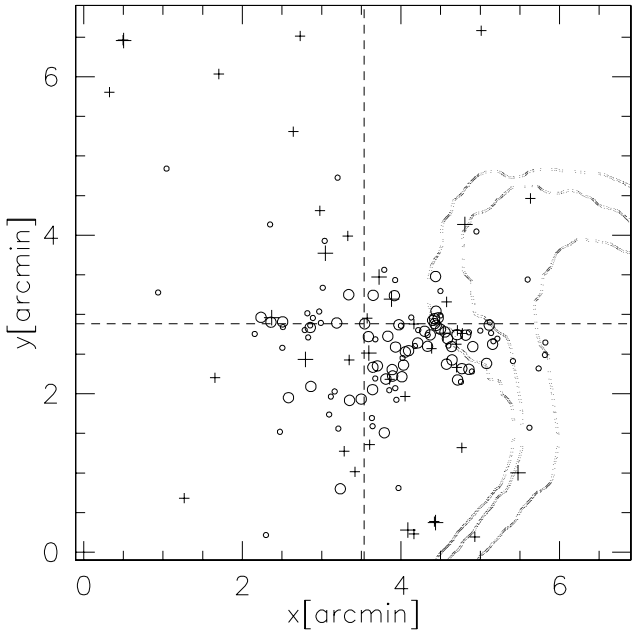}}
 \caption[]{
 Plot of the spatial distribution of the Phoenix young and intermediate age 
 populations. Open circles represent blue stars, with larger symbols  
 indicating stars brighter than $V=22.5$. Crosses are upper-AGB stars
 (a few big crosses indicate objects possibly blended). 
 We also partially reproduce the (0.5, 1, 2) $\times 
 10^{19}$  cm$^{-2}$ {\hbox{H{\sc i}}}\  column density
 contours from the maps of  Young \& Lo (\cite{youn+lo97})
 }
 \label{f_idrogeno}
 \end{figure}

The spatial distribution of young and intermediate age stars, and their 
relationship with  the neutral gas in Phoenix   (Young \& Lo 
\cite{youn+lo97}), are shown in Fig.~\ref{f_idrogeno}.
The original blue star sample (Sect.~\ref{s_cmd}) was split at 
$V=22.5$ in two subsets of stars older and younger than 
$\sim2\times10^8$\ yr (open circles; larger symbols represent the brighter,  
younger stars).  The location of the youngest stars suggests that the most 
recent star formation took place in a few central sites, while stars 
older than $\sim0.2$\ Gyr are spread over a larger area.  
The overall 
distribution of  blue stars appears  elongated in the NE--SW direction, i.e. 
roughly perpendicular to the halo of Phoenix 
({{cf.}\ } VDK91 and Fig.~\ref{f_magmap}) 
and pointing towards the {\hbox{H{\sc i}}}\ cloud.   
Since the radial velocity of Phoenix is not known, we do not really 
know if the {\hbox{H{\sc i}}}\ cloud~A is physically associated  with the galaxy, 
or the Magellanic Stream, or neither  ({{e.g.},\ } Young \& Lo \cite{youn+lo97}). 
However, if we assume that cloud~A is associated with the galaxy 
(as its location  and shell-like appearance seem to suggest), 
then the mass in  neutral gas  would be 
$1.2\times10^5$\  {$M_{\odot}$}. 
This corresponds to  
$M_{\rm H\,I}$/$L_V \approx 0.04$, an  order of magnitude  smaller  than 
the {\hbox{H{\sc i}}}\ content in dwarf irregulars, which typically have a  gas-to-star mass 
ratio of the order of unity ($M_{{\hbox{H{\sc i}}}}/L_V=0.8-3$;  Carignan {{et al.}\ } 
\cite{cari+98}). 
Among the several plausible scenarios, we only briefly comment on the possibility 
that the cloud~A  consists of  gas accumulated from mass lost by 
evolved stars, blown out by the energy input of the most recent star formation 
episode.  
We have made a simple calculation of the mass released by an old population by 
adopting 
a total mass loss rate from red giants, AGB stars and planetary nebula ejecta 
of 0.015 {$M_{\odot}$} yr$^{-1}$\  per $10^9$\ yr ({{cf.}\ } Mould {{et al.}\ } \cite{moul+90}).  
We obtain a gas return of  $\sim7\times10^5$ {$M_{\odot}$} \ after 10 Gyr of normal 
evolution of old stars in Phoenix, {i.e. } the observed amount of {\hbox{H{\sc i}}}\ could have been 
accumulated in $\sim2$\  Gyr.  By comparison, we estimate that only 
$\sim10^3$\ {$M_{\odot}$}\ of gas have been returned 
to the interstellar medium by type II 
supernovae evolved  from  massive ($>8$\ {$M_{\odot}$}) stars in the recent  burst.  
Although the ability of a dwarf galaxy  to  retain its gas depends on poorly 
known  physical parameters, such as the effects of SN explosions 
and the presence of a dark matter halo, these figures are not inconsistent
with an internal origin of the gas in Phoenix. 

\section{Summary and conclusions} 
\label{s_summa} 

We have presented a deep CCD study of  the Phoenix dwarf, a galaxy often 
regarded as  a transition case between gas-poor dwarf spheroidals and 
gas-rich dwarf irregulars. Here we summarize the  main conclusions of this study. 

The detection of the HB of Phoenix represents one of the main  results of this 
paper. We find a mean magnitude $V_{\rm HB}=23.78\pm 0.05$\ after 
correction for instrumental biases.  
Information on the HB morphology was obtained using a statistical approach.  
The  horizontal branch of Phoenix turns out to be well extended to the blue, 
although red stars are about twice as numerous. This morphology, similar to 
that of And~I, Leo~II, and Tucana has two important implications. First, it 
demonstrates the presence of a significant (if not dominant) population older 
than 10 Gyr. Second, it implies that Phoenix,  given its low metallicity, represents yet 
another (mild) example of the ``second parameter effect'' in 
dwarf galaxies.  If the origin of this effect is identified with an age difference, the 
bulk of the stellar populations in Phoenix would have to be younger (by 2--3 
Gyr) than stars in old halo globular clusters. 
We find that the spatial distribution of blue ($B-V < 0.6$) HB stars is 
significantly more extended than that of red giant stars. 
 Our result  indicates  that the early  star formation 
episode occurred in Phoenix on a larger spatial scale that subsequent  bursts. 
Horizontal branch stars have 
been  used as tracers of the spatial distribution of the oldest  
stellar populations in other dwarf galaxies (And~I, Fornax, Antlia),  
with similar results. 

Besides this old population, Phoenix has a significant intermediate age component. 
We have confirmed the presence of a small number of stars above the RGB tip, 
significantly in excess over field contamination, and  argued that these most likely 
trace the extended AGB of an intermediate age population.  Using a standard lifetime 
for the upper AGB stars, we estimate that the intermediate age populations contribute 
about  30--40\%  of the $V$\   luminosity of Phoenix.   
While we cannot establish the age of this component, there 
is some evidence that star formation declined since 8--10 Gyr ago. 
The candidate AGB stars seem to concentrate in  the inner part of the galaxy, although 
to a lesser degree than the young stars. 

Our wide photometric baseline has provided new information on the  young stellar 
population  in Phoenix. We have shown that the recent star formation episode,  
responsible for the sprinkling of blue stars characteristic of  this dwarf, started at least 
0.6 Gyr ago. The recent burst of stars formation
 ($\sim1-2.5\times10^8$\ yr   ago)  accounts for less than 6\% of the $V$\ luminosity 
of Phoenix and 0.2\% in terms of mass. 
The blue stars which trace  the most recent burst are concentrated in clumps or 
``associations'' near the galaxy  center, with a spatial distribution elongated in a 
direction  perpendicular  to the major axis defined by the diffuse galaxy light, and 
slightly offset towards the {\hbox{H{\sc i}}}\ cloud observed by Young \& Lo (\cite{youn+lo97}). 
The neutral gas could have been blown out  by the recent burst, 
a possibility the 
should be further investigated  when the hypothesis of a  physical 
link receives support by measurements of  the galaxy radial velocity. 

Excluding the regions of recent star formation, we have defined a clean 
sample of RGB stars that has been employed to re-derive the galaxy 
basic properties. A new distance modulus  
$(m-M)_{\rm RGB,0}=23.04 \pm 0.07$\  was  obtained using the 
well defined cutoff  of the red giant branch  in the $I$, $(V-I)$\ diagram. 
More importantly, we have obtained for the first time an independent 
estimate of the distance to Phoenix, 
$(m-M)_{\rm HB,0}=23.21 \pm 0.08$, 
based on the mean level of horizontal branch stars. 
 The mean of the two independent measurements gives a distance 
$(m-M)_0=23.12 \pm 0.08$, which confirms previous estimates. 
A mean metal abundance [Fe/H]$=-1.81 \pm 0.10$ was obtained with a  direct 
comparison of  the upper part of the RGB with the fiducial sequences  of  template 
Galactic globular clusters. 
A careful analysis based on extensive artificial star tests and 
comparison with previous photometry  confirms the presence of  an intrinsic color 
scatter in the red giants of  Phoenix, 
$\sigma_{(V-I)}=0.08\pm0.01$\ mag, corresponding  to a metallicity 
dispersion  $\sigma_{\rm [Fe/H]}=0.23\pm0.03$\ dex.  

In conclusion, this paper provides new evidence  that Phoenix has had an 
extended history of star formation. Its stellar populations appear fundamentally similar 
to those found in many dwarf spheroidal galaxies -- even the young population has 
ages comparable to those of the youngest stars in Fornax (Beauchamp {{et al.}\ } 
\cite{beau+95}, Stetson {{et al.}\ } \cite{stet+98}, Saviane \& Held 
\cite{savi+held98}; Held et al. \cite{held+99}). Its {\hbox{H{\sc i}}}\ content (if any) is comparable 
to the amount of neutral gas found in Sculptor. 
In view of these results, we are inclined to regard Phoenix as a low-mass dwarf 
spheroidal seen in the middle of a star formation  episode (during which gas is perhaps 
expelled). 
This similarity in the stellar populations of Phoenix and more luminous 
dSph's is noteworthy given the difference either in mass 
or in location with respect to the 
big Local Group spirals. There appears to be no obvious correlation between  
the timescale of star formation and galaxy mass, although there is one between 
mass and production of heavy elements, as made evident  by the well-know  
luminosity-metallicity correlation ({{e.g.},\ } Buonanno et al. \cite{buon+85}). 
The effects of the environment are not clear either, although Phoenix seems 
to fit well the trend suggested by van den Bergh (\cite{vdbe94}) between the 
presence of  young or intermediate age populations in dSph's  and their 
distance from the Galaxy or M31. 
The origin of the striking difference in the star formation histories of Phoenix and 
Tucana (both are isolated Local Group dwarfs) remains rather puzzling, since these 
two dwarfs not only have comparable luminosities and metal abundances, but also a 
similar HB morphology, which probably indicates an old population formed nearly at 
the same epoch. 

\acknowledgements{
It is a pleasure to thank C.~Maraston and G.~P.~Bertelli for clarifying 
discussions on stellar population synthesis in metal-poor systems.  
We are grateful to C. Chiosi and S. Ortolani for careful reading of 
the original manuscript, and to the referee, Dr.~Demers, for useful  
comments and suggestions. 
G. Da Costa kindly provided unpublished information on his
spectroscopic observations of candidate AGB stars. 
I.S. acknowledges support of ANTARES, an astrophysics
 network funded by the HCM programme of the European Community.
Y.~M. acknowledges support from the Italian Ministry of Foreign 
Affairs and the Dottorato di Ricerca program at the University of Padova. 
}

\end{document}